\documentclass[conference,onecolumn]{IEEEtran}
\IEEEoverridecommandlockouts
\usepackage{cite}
\usepackage{amsmath,amssymb,amsfonts}
\usepackage{algorithmic}
\usepackage{graphicx}
\usepackage{textcomp}
\usepackage{caption}
\usepackage{multirow}
\usepackage{graphicx}
\usepackage{subcaption}

\usepackage[ruled,linesnumbered]{algorithm2e}
\usepackage{xcolor}
\def\BibTeX{{\rm B\kern-.05em{\sc i\kern-.025em b}\kern-.08em
    T\kern-.1667em\lower.7ex\hbox{E}\kern-.125emX}}
\begin{document}

\title{Intelligent Agricultural Greenhouse Control System Based on Internet of Things and Machine Learning\\
}

\author{ Cangqing Wang$^1*$, Jiangchuan Gong$^1$
\thanks{$^1*$Cangqing Wang be with Boston University in Boston, MA02215. Correspondence to Cangqing Wang via email: {\tt\small \{kriswang\}@bu.edu}}
\thanks{$^1$Jiangchuan Gong is an independent researcher. Correspondence to Jiangchuan Gong via email: {\tt\small \{jackgong151823\}@gmail.com}}
}

\maketitle

\begin{abstract}This study endeavors to conceptualize and execute a sophisticated agricultural greenhouse control system grounded in the amalgamation of the Internet of Things (IoT) and machine learning. Through meticulous monitoring of intrinsic environmental parameters within the greenhouse and the integration of machine learning algorithms, the conditions within the greenhouse are aptly modulated. The envisaged outcome is an enhancement in crop growth efficiency and yield, accompanied by a reduction in resource wastage. In the backdrop of escalating global population figures and the escalating exigencies of climate change, agriculture confronts unprecedented challenges. Conventional agricultural paradigms have proven inadequate in addressing the imperatives of food safety and production efficiency. Against this backdrop, greenhouse agriculture emerges as a viable solution, proffering a controlled milieu for crop cultivation to augment yields, refine quality, and diminish reliance on natural resources \cite{ritchie2023co₂}. Nevertheless, greenhouse agriculture contends with a gamut of challenges. Traditional greenhouse management strategies, often grounded in experiential knowledge and predefined rules, lack targeted personalized regulation, thereby resulting in resource inefficiencies. The exigencies of real-time monitoring and precise control of the greenhouse's internal environment gain paramount importance with the burgeoning scale of agriculture. To redress this challenge, the study introduces IoT technology and machine learning algorithms into greenhouse agriculture, aspiring to institute an intelligent agricultural greenhouse control system conducive to augmenting the efficiency and sustainability of agricultural production. 
\end{abstract}


\section{Introduction}
\label{sec:introduction}
In the formulation of the intelligent agricultural greenhouse control system, ethical considerations and system comprehensiveness demand meticulous attention. The real-time monitoring of greenhouse environmental parameters necessitates the collection and processing of copious sensitive data, thereby engendering privacy concerns for agricultural stakeholders, greenhouse equipment manufacturers, and other involved parties\cite{jeffry2021greenhouse}. In response, the study employs secure data transmission and storage mechanisms that encrypt sensitive information, ensuring exclusive access to critical data by authorized personnel. Additionally, a well-defined data use policy delineates ownership and usage rights, complemented by data desensitization measures to mitigate potential privacy risks.

The advent of IoT and machine learning may precipitate a paradigm shift in agricultural management, potentially engendering technological adaptation and employment challenges for practitioners. To mitigate this, the study proffers training and support initiatives, ensuring that agricultural practitioners are adept at harnessing new technologies. Concurrently, advocacy for open communication channels between practitioners, scientists, and governments is underscored to deliberate on the impact of technological integration on traditional agricultural practices.

Post-implementation of smart technologies, ensuring the sustainability and enduring maintenance of the system becomes a pertinent concern. To address this, the system is meticulously designed with modularity and upgradability, allowing for facile updates to hardware and software. Furthermore, a judicious maintenance plan and support system are instituted to uphold the system's functionality over the long term.

The introduction of novel technologies harbors the prospect of an uneven distribution of resources within society, potentially depriving certain regions or agricultural producers of the benefits of smart agriculture. To safeguard social equity and accessibility, the study formulates policies and programs to ensure the equitable and accessible deployment of new technologies across all societal strata. This encompasses the provision of technical support and training in remote areas to bridge the digital divide.

The study's objective revolves around the design and implementation of an intelligent agricultural greenhouse control system predicated on the IoT and machine learning, with the overarching aim of enhancing crop growth efficiency and yield\cite{fang2021survey}. By scrutinizing internal environmental parameters such as temperature, humidity, and light, coupled with the judicious application of machine learning algorithms, real-time and precise control of these parameters is sought. The anticipated outcome is an adaptive system enabling farmers to cater to diverse crop needs, mitigate resource wastage, and elevate production efficiency.

The significance of this research is multifaceted. Primarily, the implementation of an intelligent agricultural greenhouse control system is poised to usher in technological innovation in agricultural production, fostering the metamorphosis from traditional to intelligent and precision agriculture. Additionally, through curtailed energy, water, and fertilizer usage, the system is poised to be a linchpin in supporting sustainable agricultural development. Most crucially, the augmentation of agricultural production contributes tangibly to meeting the escalating global demand for food, thereby contributing to societal well-being and sustainable economic development.

The key trajectory for future development hinges upon the assimilation of emerging technologies, notably the incorporation of advanced sensor technologies such as high-resolution multi-spectral sensors. These sensors are envisaged to yield more comprehensive environmental data, coupled with the infusion of advanced wireless communication technologies to ensure expeditious and reliable data transmission. This integration, however, may grapple with challenges related to hardware compatibility and data processing capabilities. The envisaged solution entails the adoption of a modular design facilitating the facile replacement and upgrade of sensors, coupled with the optimization of data processing algorithms to accommodate expansive datasets.

An additional focal point for future development is the perpetual refinement of machine learning algorithms. This entails the optimization of extant algorithms, the integration of deep learning methodologies to fortify the modeling of intricate relationships within greenhouse environments, and the infusion of online learning techniques to facilitate dynamic adaptation to changing agricultural terrains. The introduction of deep learning algorithms, however, may necessitate augmented computing resources and mandates addressing energy consumption and hardware requisites. Simultaneously, the challenge inherent in online learning technology involves real-time model updates without compromising system performance, necessitating consideration of algorithmic stability and real-time efficacy.

Future developmental strides should be directed towards the seamless integration of the intelligent agricultural greenhouse control system with the entire agricultural ecosystem. This encompasses interoperability with field management and supply chain systems, thereby actualizing a comprehensive agricultural intelligence framework. However, the assimilation of diverse systems may implicate issues of data standardization, warranting the establishment of industry standards to foster the harmonious development of disparate systems.

Critical to the realization of future developments is the amplification of farmers' acceptance of smart farming technologies. Through systematic training and education initiatives, the study aims to facilitate a nuanced understanding among farmers, enabling them to adeptly operate intelligent greenhouse control systems. Recognizing the potential influence of cultural and geographical disparities on social acceptance, the study underscores the establishment of regular training programs, the leveraging of community resources, and the augmentation of farmers' awareness and trust in new technologies as pivotal strategies to overcome these challenges. Through concerted efforts to address these challenges and persisting in propelling technological innovation, smart agricultural greenhouse control systems are poised to assume an expanded role in propelling the sustainable advancement of agriculture and overall societal progress.

Recent years have witnessed remarkable strides in the application of IoT and machine learning technologies in agriculture. IoT technology facilitates real-time connectivity of agricultural equipment and the collection of extensive environmental data. These data serve as a fertile information bedrock for machine learning algorithms, enabling intelligent management of agricultural production processes through data analysis and model training. 

Within agriculture, machine learning algorithms find widespread utility in forecasting, optimization, and decision support. Historical data analysis empowers the construction of models predicting crop growth under diverse meteorological conditions, thereby enabling precise regulation of growth environments. Furthermore, machine learning can dynamically adjust greenhouse conditions based on real-time data, aligning with distinct stages of crop growth to amplify production efficiency\cite{chen2023soc}. 

A comprehensive scrutiny of extant literature underscores the prodigious potential of IoT and machine learning in agriculture. However, for the exhaustive exploration and practical implementation of an intelligent agricultural greenhouse control system, myriad challenges and issues persist. Consequently, this study aspires to bridge extant research lacunae, proffering an innovative intelligent greenhouse control system and validating its efficacy in enhancing agricultural production efficiency and sustainability through real-world case studies. 

Within current literature and practical applications, this study focalizes on rectifying specific deficiencies and research gaps in the domain of intelligent agricultural greenhouse control systems, thus fostering intelligent and sustainable agricultural development. Foremost among these endeavors is the amelioration of real-time interconnection and data collection integration, achieved through IoT technology to enable real-time connectivity of agricultural equipment and the aggregation of voluminous environmental data. This robust data infrastructure forms the bedrock for the design of an intelligent agricultural greenhouse control system adept at processing and leveraging diverse environmental data efficiently.

The study keenly addresses the lacunae inherent in intelligent decision support systems, acknowledging that the application of machine learning algorithms in agriculture necessitates improvements in accuracy and real-time capabilities. Through the infusion of innovative machine learning algorithms, the research endeavors to fortify the intelligent management of agricultural production processes, thereby facilitating more precise and real-time decision-making in agriculture.

A discernible focus of the study is the resolution of the comprehensive optimization challenge in greenhouse environments. The recognition of a research gap in the extant literature pertaining to the holistic optimization of greenhouse environmental parameters propels the study to center on comprehensive greenhouse environment optimization, including temperature, humidity, light, and other facets\cite{domb2022satellite}. The overarching objective is to fashion a system that comprehensively meets the diverse needs of crop growth across distinct stages.

The study accentuates the dearth of comprehensive research and practical application, underscoring the inadequacy in the existing body of research on the IoT and machine learning in agriculture. By propounding innovative intelligent greenhouse control systems and substantiating their efficacy through practical case studies, the study aspires to rectify this research gap, thereby offering more robust theoretical and practical support in the domain of intelligent agriculture. 

\section{Greenhouse internal environmental monitoring system }

\begin{figure}[htbp]
\centerline{\includegraphics[height=6cm]{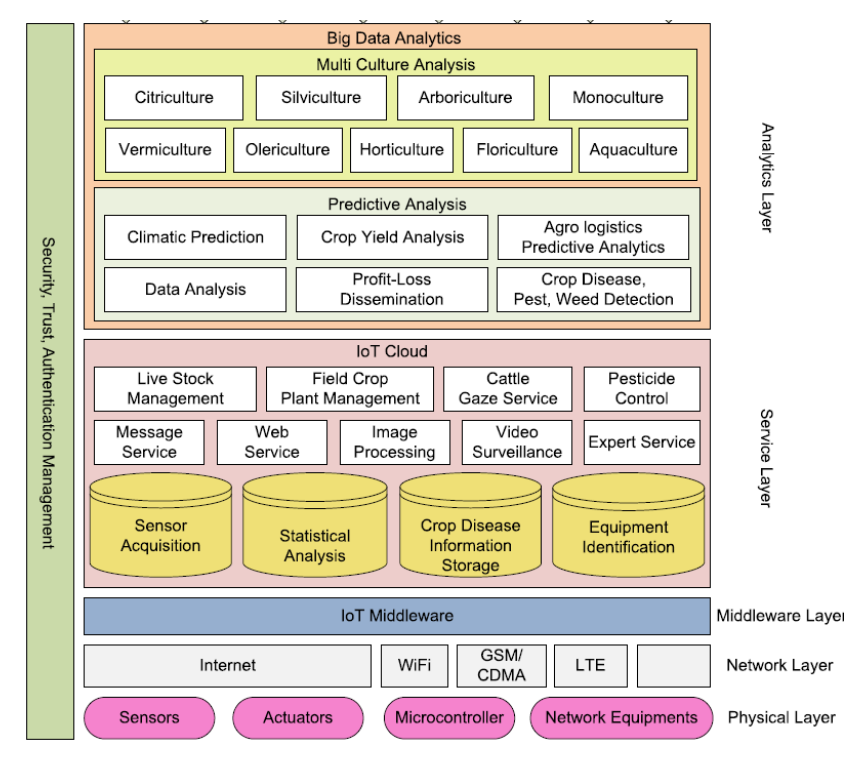}}
\caption{The structure diagram of the intelligent agricultural temperature control system of the Internet of Things }
\label{fig-1}
\end{figure}

\subsection{Sensor selection and layout: temperature, humidity, light and other parameters }
Achieving precision in monitoring the internal environment of a greenhouse necessitates the judicious selection of sensors and their strategic deployment. This ensures a comprehensive and accurate acquisition of environmental parameters across various zones within the greenhouse.

\subsubsection{Sensor Selection }
In the realm of greenhouse environment monitoring, pivotal factors influencing crop growth encompass temperature, humidity, and light. Consequently, we opted for temperature sensors, humidity sensors, and light sensors for surveillance. Digital temperature sensors such as DS18B20, DHT series sensors for humidity, and photoresistors or photodiodes for illuminance were employed. 

\begin{figure}[htbp]
\centerline{\includegraphics[height=6cm]{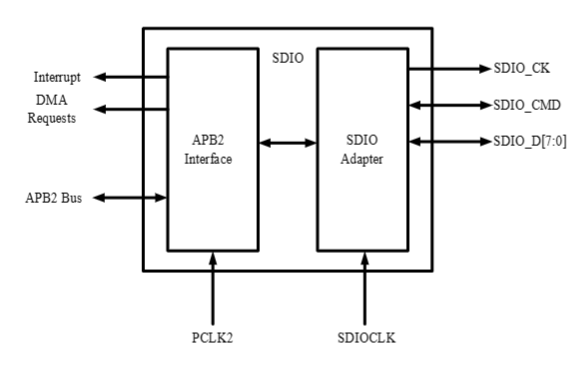}}
\caption{Circuit diagram of light source module }
\label{fig-2}
\end{figure}

\subsubsection{Sensor Layout }The efficacy of the monitoring system hinges on the meticulous arrangement of sensors. We evenly distribute temperature and humidity sensors throughout diverse areas of the greenhouse to obtain comprehensive environmental data \cite{hoseinzadeh2024ai}. The light sensor is strategically positioned to discern changes in light, enabling real-time monitoring of light conditions. 

\subsection{Data acquisition and transmission: real-time monitoring of the greenhouse internal environment }Real-time monitoring of the greenhouse internal environment necessitates the seamless collection and transmission of sensor data. Leveraging IoT technology, acquired data is transferred to the cloud for processing and storage.
\subsubsection{Data Collection}Microcontrollers such as Arduino or Raspberry Pi facilitate data collection by greenhouse sensors. Temperature (T), humidity (H), illuminance (L), and other environmental parameters are collected and converted into digital signals through an analog-to-digital converter (ADC) for subsequent processing. 

\subsubsection{Data Transfer}
Wireless communication modules like Wi-Fi or LoRa are employed to transmit collected data to the cloud. Cloud servers receive and store the data, offering real-time accessibility. This monitoring system ensures a timely comprehension of changes in the internal greenhouse environment, providing a foundation for subsequent decision-making and regulation \cite{maraveas2022incorporating}. 

\begin{table}[h]
\centering
\caption{Instructions and Their Functions}
\label{table:instructions-functions}
\begin{tabular}{|p{100pt}|p{125pt}|}
\hline
\textbf{Instructions} & \textbf{Functions} \\
\hline
AT+CWMODE=1 & Set the WiFi module to STA mode \\
\hline
AT+RST & Restart the WiFi module to take effect \\
\hline
AT+CWJAP="xx","xxxxxxxx" & Connect to a WiFi hotspot with SSID "xx" and password "xxxxxxxx" \\
\hline
AT+CIPMUX=0 & Enable single connection mode \\
\hline
AT+CIPSTART="TCP", "192.168.1.XXX", 8000 & Establish a TCP connection to IP 192.168.1.XXX on port 8000 \\
\hline
AT+CIPMODE=1 & Enable transparent transmission mode \\
\hline
AT+CIPSEND & Begin data transmission \\
\hline
\end{tabular}
\end{table}

\subsection{Data processing and storage: Cloud computing applications}

In the cloud, cloud computing technology is utilized for processing and storing collected data. Processing involves cleaning, normalization, and denoising to guarantee data quality and accuracy. Additionally, machine learning algorithms analyze historical data, facilitating model training to support subsequent prediction and real-time regulation of greenhouse conditions.

\subsubsection{Data processing}

During data processing, outlier detection and repair are conducted to ensure data reliability. The cleaned data is normalized, unifying sensor data scales for convenient algorithmic processing.

\paragraph{Anomaly Detection and Repair During Data Processing\\}

\textbf{1. Outlier detection:}

In the intelligent agricultural greenhouse control system, data may be subject to various disturbances, such as sensor failures or sudden environmental changes, leading to outlier generation. Ensuring data reliability requires the detection and repair of outliers.

\textbf{1.1 Standard deviation method:}

A common approach to outlier detection is the standard deviation method. For each sensor's collected data, the deviation from the average is calculated, and data exceeding the preset threshold are treated as outliers. The formula is as follows:

\begin{equation}
\text{Deviation} = \frac{\text{Data Point} - \text{Mean}}{\text{Standard Deviation}}
\end{equation}

If the deviation exceeds a set threshold, the data point is marked as an outlier.

\textbf{1.2 Box diagram method:} An alternative technique for identifying outliers involves the application of the boxplot method. This approach leverages the interquartile range (IQR) to pinpoint data points exceeding predefined upper and lower thresholds, thus classifying them as outliers. The determination of these limits involves the computation of the quartiles, where the upper limit (UL) is expressed as \(Q3 + 1.5 \times IQR\), and the lower limit (LL) is given by \(Q1 - 1.5 \times IQR\).

Here, \(Q1\) and \(Q3\) signify the first and third quartiles, respectively, and \(IQR\) represents the interquartile range.

\textbf{2. Data normalization:}
Data standardization involves the transformation of sensor data onto a standardized scale, ensuring homogeneity and streamlining subsequent algorithmic processing. This practice is instrumental in enhancing algorithm performance and mitigating undesirable impacts arising from variations in data scales. 

\textbf{2.1 Min-Max normalization: }Min-Max normalization is a prevalent approach for linearly mapping data to a scale within the range of 0 to 1. The specific formula is articulated as follows: 
\begin{equation}
x_{\text{normalized}} = \frac{x - x_{\text{min}}}{x_{\text{max}} - x_{\text{min}}}
\end{equation}
This method ensures that each sensor's data is uniformly scaled, contributing to consistent weighting during algorithmic analysis. 

\textbf{2.2 Z-Score normalization: }Z-Score normalization transforms data into a standard normal distribution with a mean of 0 and a standard deviation of 1 by calculating the standard score. The formula is expressed as:
\begin{equation}
    z = \frac{x - \mu}{\sigma}
\end{equation}
Z-Score normalization is particularly applicable in scenarios where data approximate a normal distribution, effectively reducing the impact of outliers on the normalized outcomes. 

\subsubsection{Application of machine learning algorithm}

Machine learning algorithms are employed for data analysis and model training, particularly in the context of the dynamic and temporal changes within a greenhouse environment. Recurrent neural networks, for instance, can be utilized for predicting greenhouse conditions. The fundamental formula governing such networks is presented as:

\begin{equation}
h_t = f(W \cdot h_{t-1} + U \cdot x_t + b)
\end{equation}

Where \(h_t\) denotes the hidden state at time \(t\), \(f\) represents the activation function, \(h_{t-1}\) signifies the hidden state at time \(t-1\), \(x_t\) represents the input at time \(t\), \(W\) and \(U\) are weight matrices, and \(b\) is the bias.

\section{Application of machine learning algorithms }
\subsection{Data analysis and model training: The use of historical data }
In the context of intelligent agricultural greenhouse control systems, the application of machine learning algorithms holds significant importance. Through the analysis of historical data and subsequent model training, the system acquires the ability to discern intricate relationships among greenhouse environmental parameters, thereby enabling accurate predictions of future conditions. 
\subsubsection{Data Preparation}
The archival data encompass the intrinsic environmental parameters within the greenhouse, namely temperature, humidity, and illuminance, alongside their corresponding regulatory actions involving adjustments to the greenhouse conditions. A prerequisite is the preprocessing of this data, with a temporal series data format chosen to enable the model's comprehensive utilization of temporal information \cite{fang2021survey}. 
\subsubsection{Feature engineering }Preceding model training, a crucial step involves feature engineering and the judicious selection of pertinent feature variables. Notably, features such as temperature, humidity, and illuminance over a temporal span serve as input variables, while future temporal points' environmental parameters serve as output labels\cite{ullah2022design}.
\subsubsection{Model selection}

Given the sequential and dynamic nature of greenhouse environmental changes, the adoption of a Recurrent Neural Network (RNN) for model training is deemed appropriate\cite{gong2021deep}. RNN, a deep learning model tailored for sequential data, exhibits a fundamental structure as expressed below:

\begin{align}
h_t = f(H_{t-1} , X_{t}) \\
y_t = g(H_{y})
\end{align}

Where:
\begin{itemize}
    \item \(h_t\) signifies the hidden state at time \(t\).
    \item \(f\) and \(g\) represent the activation and output functions, respectively.
    \item \(h_{t-1}\) denotes the hidden state of time \(t-1\).
    \item \(x_t\) constitutes the input at time \(t\).
    \item \(y_t\) is the output at time \(t\).
    \item \(W_{hh}\), \(W_{hx}\), and \(W_{yh}\) are weight matrices.
    \item \(b_h\) and \(b_y\) are bias terms for the hidden and output layers, respectively.
\end{itemize}

During the intelligent agricultural greenhouse control system's model training, myriad challenges are addressed through targeted countermeasures. Imbalances in historical data necessitate the application of oversampling or undersampling techniques to rectify scenario-specific sample scarcity. Additionally, weight adjustment mitigates learning disparities across categories, enhancing the model's proficiency for select classes.

Convergence issues may arise during model training, particularly due to the sequential and dynamic nature of the greenhouse environment. Addressing this, learning rate adjustments prove effective in modulating the equilibrium model's learning pace. Furthermore, introducing more intricate models, such as augmenting the layers or units of recurrent neural networks (RNNs), enhances the model's fitting capacity.

Complex models processing time series data may encounter overfitting concerns. To counteract this, regularization techniques such as L1 and L2 regularization are applied to curb overfitting by penalizing model complexity. Simultaneously, Dropout technology is integrated to randomly deactivate certain neurons, reducing the risk of neural network overfitting.

The pivotal roles of data preprocessing and feature engineering in model performance underscore the importance of meticulous feature selection related to dynamic changes in the greenhouse environment, thereby augmenting the model's generalization capabilities\cite{liu2022long}. Normalization and standardization of data ensure uniform feature value ranges, thereby facilitating the stable training of the model. These adaptive methodologies collectively aim to effectively surmount diverse challenges encountered in model training, thereby enhancing the performance and reliability of the intelligent agricultural greenhouse control system.

\subsection{Prediction of greenhouse conditions: a model based on machine learning }
Following model training, the predictive capacity of the model is harnessed to anticipate greenhouse environmental parameters. By inputting present-moment environmental data, the model extrapolates greenhouse conditions at specified future time points. 

\subsection{Overview of control strategies and algorithms of real-time regulation system: }
Within the real-time control system, precision in regulating greenhouse conditions based on model output is achieved through the employment of specific control strategies and algorithms in this investigation.

\begin{figure}[htbp]
\centerline{\includegraphics[height=6cm]{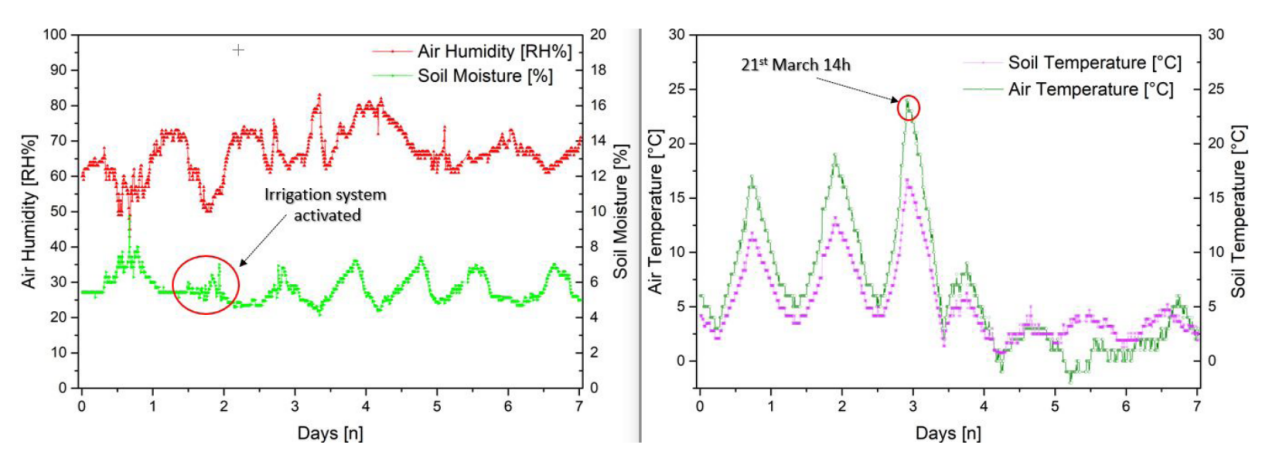}}
\caption{Measurement results using the smart agricultural system (Air Humidity, Soil Moisture (left)and Soil Temperatire, Air Temperature (right))}
\label{fig-3}
\end{figure}

\begin{figure}[htbp]
\centerline{\includegraphics[height=6cm]{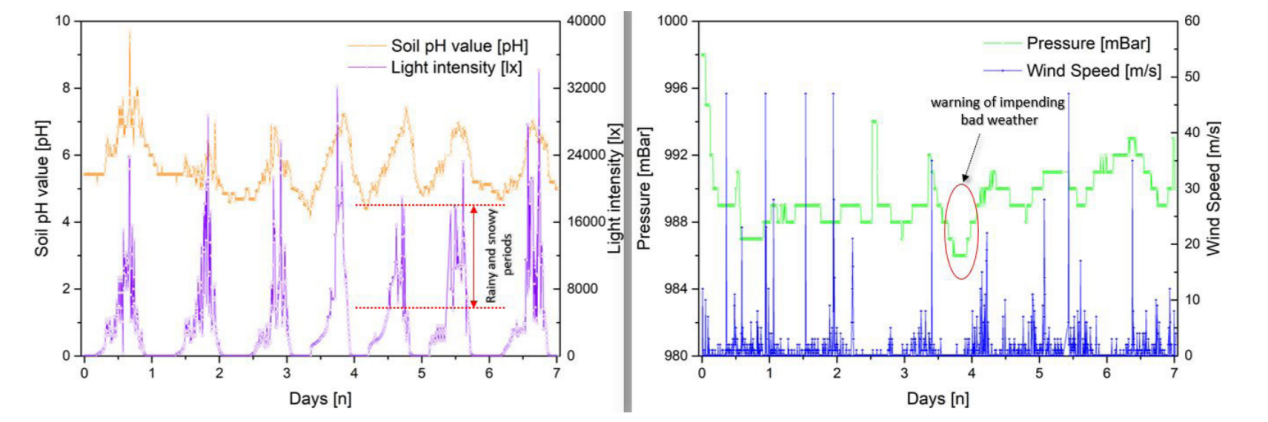}}
\caption{Measurement results using the smart agricultural system (Soil pH value, Light Intensity (left) and Air Pressure, Wind speed (right))}
\label{fig-4}
\end{figure}

\subsubsection{PID Controller}This system incorporates the classical Proportional Integral Derivative (PID) controller as its foundational control algorithm. The PID controller computes control output by considering the current error (the deviation between actual and target values), the integral of the error, and the error's rate of change. Applied distinctly to temperature, humidity, and irrigation parameters, the PID controller, through adjustments in proportional coefficient, integration time, and differential time, accomplishes precise environmental condition modifications. 

\subsubsection{Model Predictive Control (MPC)}Model Predictive Control (MPC), an advanced method founded on system dynamic models, utilizes model predictions for a future timeframe as control input to optimize greenhouse environmental control strategies. Acknowledging system time-variability and nonlinearity, MPC enhances responsiveness to diverse meteorological conditions and crop growth stages, augmenting adaptability and regulation robustness. 

\subsubsection{Fuzzy control system}The fuzzy control system, rooted in fuzzy set theory, is suited for intricate and nonlinear systems. In greenhouse environmental control, fuzzy control maps fuzzy natural language rules to tangible control actions. For instance, intelligent greenhouse regulation is achieved by articulating the fuzzy rule "if the predicted temperature rises and humidity is low, increase irrigation."
\subsubsection{Integrated Control Policy}
To leverage various control algorithms' strengths, the system adopts an integrated control strategy. Adapted to real-time circumstances, the system dynamically switches or combines PID controller, MPC, and fuzzy control system to effect multi-level and multi-faceted regulation of diverse environmental parameters, thereby enhancing overall system performance. 

Through comprehensive application of the aforementioned control strategies and algorithms, the real-time control system exhibits heightened flexibility and intelligence in responding to diverse environmental changes and model predictions. This ensures the greenhouse's internal environment consistently aligns with optimal conditions, thereby improving crop growth efficiency and yield. 

\subsubsection{ Impact of Real-Time Regulation}The real-time regulatory efficacy is scrutinized through continuous monitoring and control measures, elucidating the dynamic fluctuations of internal environmental parameters within the anticipated range within the greenhouse. This observation signifies the adept responsiveness of the real-time control system to the predictive outcomes derived from the model, thereby ensuring the maintenance of an optimal growth environment. 

\begin{table}[!ht]
\centering
\caption{\textbf{Units for Magnetic Properties}}
\label{table}
\setlength{\tabcolsep}{3pt}
\begin{tabular}{|p{30pt}|p{30pt}|p{30pt}|p{35pt}|p{35pt}|p{40pt}|}
\hline
Number & Air Temperature & Air Humidity \% & Soil Temperature & Soil Moisture \% & CO2 Concentration (PPM) \\ \hline
1 & 21.3 C & 29.1 & 25.5 & 3000.1 & 441 \\ \hline
2 & 21.6 C & 29.1 & 25.6 & 3000.2 & 444 \\ \hline
3 & 21.6 C & 29.3 & 25.7 & 3000.2 & 449 \\ \hline
4 & 21.7 C & 30.0 & 25.8 & 2999.2 & 455 \\ \hline
5 & 21.9 C & 30.1 & 26.9 & 2999.1 & 459 \\ \hline
6 & 21.9 C & 30.1 & 27.1 & 3000.2 & 467 \\ \hline
7 & 22.1 C & 29.1 & 29.5 & 3001.2 & 471 \\ \hline
8 & 22.2 C & 29.2 & 27.8 & 3005.1 & 473 \\ \hline
9 & 23.2 C & 30.1 & 28.2 & 3010.1 & 482 \\ \hline
10 & 23.5 C & 30.0 & 28.9 & 3011.2 & 491 \\ \hline

\end{tabular}
\label{tab2}
\end{table}

\begin{table}[!ht]
\centering
\caption{The testing results of the Dedicated testing tools }
\label{table:environmental-parameters}
\setlength{\tabcolsep}{3pt}
\begin{tabular}{|p{30pt}|p{30pt}|p{30pt}|p{35pt}|p{35pt}|p{40pt}|}
\hline
Number & Air Temperature & Air Humidity \% & Soil Temperature & Soil Moisture \% & CO2 Concentration (PPM) \\ \hline
1 & 20.3 C & 28.1 & 25.2 & 2999.1 & 439 \\ \hline
2 & 20.5 C & 28.1 & 25.2 & 2995.2 & 442 \\ \hline
3 & 20.6 C & 28.2 & 25.3 & 2991.2 & 448 \\ \hline
4 & 20.5 C & 29.0 & 25.5 & 2988.2 & 452 \\ \hline
5 & 20.6 C & 29.0 & 26.1 & 2995.1 & 458 \\ \hline
6 & 20.7 C & 29.0 & 26.5 & 2999.2 & 465 \\ \hline
7 & 21.3 C & 28.0 & 26.9 & 3000.2 & 469 \\ \hline
8 & 21.8 C & 29.0 & 27.1 & 3004.1 & 472 \\ \hline
9 & 22.1 C & 29.2 & 27.9 & 3009.1 & 480 \\ \hline
10 & 22.5 C & 29.3 & 28.2 & 3010.2 & 489 \\ \hline
\end{tabular}
\label{tab3}
\end{table}
Through the application of such machine learning algorithms, we realize the intelligent management of the greenhouse environment, improve the efficiency of crop growth, and reduce the waste of resources. In practical applications, such a system can provide more reliable and efficient support for agricultural production and promote the development of agriculture in a smart and sustainable direction. 

\section{Improvement of crop growth efficiency }
\subsection{ Optimization of the growing environment: according to the needs of different crops }Different crops have their own unique needs for the growing environment, including temperature, humidity, light and other factors. Through the intelligent agricultural greenhouse control system, we are able to optimize the growing environment according to the needs of different crops, thereby improving their growth efficiency. 
\subsubsection{Adjustment of growth environment parameters }Based on the output of the machine learning model, parameters such as temperature, humidity, and illuminance in the greenhouse are dynamically adjusted. Taking temperature as an example, we use a feedback control algorithm to calculate the demand for heating or cooling in the greenhouse by measuring the difference between the current temperature and the target temperature, and adjust the temperature through the control equipment. The basic formula is as follows:

\begin{equation}
    \text{Adjustment amount} = K_p \cdot e + K_i \cdot \int e \, dt + K_d \cdot \frac{de}{dt}
\end{equation}

Where:
\begin{itemize}
    \item \(K_p\), \(K_i\), and \(K_d\) are the proportional, integral, and derivative gain parameters, respectively.
    \item \(e\) represents the error term.
    \item \(\int e \, dt\) represents the integral of the error over time.
    \item \(\frac{de}{dt}\) represents the derivative of the error with respect to time.
\end{itemize}

\subsubsection{Individual regulation of different crops }Through the prediction and regulation of greenhouse environment, we can realize the individualized growth environment adjustment of different crops. For example, for tropical fruit crops, we can provide higher temperature and humidity, and for vegetable crops adapted to cold climates, we can adjust the temperature accordingly. 

\subsection{Improvement of resource utilization efficiency }While improving crop yields, the smart agricultural greenhouse control system also aims to optimize the efficiency of resource use, including energy and water resources. 

\subsubsection{Energy consumption optimization}
Through real-time monitoring and control systems, we can precisely control the temperature inside the greenhouse and reduce energy waste. For example, based on external weather forecasts and real-time data on the greenhouse's internal environment, the system can intelligently adjust the use of heating and ventilation equipment to achieve efficient use of energy.

\subsubsection{Reduce fertilizer and water use }Based on the greenhouse environmental parameters predicted by the model, the system precisely controls the amount of irrigation and fertilizer application to reduce the use of water resources and fertilizers. By adjusting the amount of irrigation water at different growth stages, the system is able to meet the needs of the plants and avoid the waste of resources caused by over-irrigation. 

\section{Energy saving and environmental protection considerations }
\subsection{Optimization of energy consumption}The intelligent agricultural greenhouse control system can optimize energy use and reduce energy consumption through real-time monitoring of greenhouse environmental parameters and prediction of machine learning algorithms. 

\subsubsection{Energy consumption model}

We built an energy consumption model, taking into account factors such as temperature difference, light intensity, wind speed, etc., to predict the heating and ventilation needs of the greenhouse at the current moment. The basic formula of the model is as follows:

\begin{equation}
    \text{Energy consumption} = K_1 \cdot \Delta T + K_2 \cdot L + K_3 \cdot W
\end{equation}

Where:
\begin{itemize}
    \item \(K_1\), \(K_2\), and \(K_3\) are the adjustment parameters.
    \item \(\Delta T\) represents the temperature difference (indoor and outdoor temperature difference).
    \item \(L\) represents light intensity (lighting conditions).
    \item \(W\) represents wind speed (wind intensity).
\end{itemize}

\subsubsection{Control Policy}

Through the prediction of the energy consumption model, the system can develop intelligent control strategies. For example, in the case of higher external temperatures predicted, the system can reduce the temperature inside the greenhouse in advance, reducing the cooling load and thereby reducing energy consumption.

\subsection{Reduce fertilizer and water use}

Through precise control of irrigation and fertilization, the smart agricultural greenhouse control system aims to reduce the overuse of fertilizers and water resources, thereby reducing the burden of agriculture on the environment.

\subsubsection{Water resource utilization model}

By establishing a water resource utilization model, the system considers factors such as the internal environment of the greenhouse, soil moisture, and crop water demand to predict the irrigation demand at the current moment. The basic formula of the model is as follows:

\begin{equation}
 I = K_4 \cdot \Delta M + K_5 \cdot C   
\end{equation}

Where:
\begin{itemize}
    \item \(I\) represents the irrigation quantity.
    \item \(K_4\) and \(K_5\) are the adjustment parameters.
    \item \(\Delta M\) represents the soil moisture difference (difference between the current soil moisture and the target moisture).
    \item \(C\) represents the crop water requirement (current crop growth water requirement).
\end{itemize}

\subsubsection{Fertilization optimization algorithm}

In terms of fertilization, the system reduces the excessive use of fertilizer by establishing a fertilization optimization algorithm according to the growth stage and nutrient requirements of crops. The basic formula of the algorithm is as follows:

\begin{equation}
    F = K_6 \cdot G + K_7 \cdot N
\end{equation}

Where:
\begin{itemize}
    \item \(F\) represents the rate of fertilizer application.
    \item \(K_6\) and \(K_7\) are the adjustment parameters.
    \item \(G\) represents the crop growth stage (current crop growth state).
    \item \(N\) represents the soil nutrient content (nutrient level in the soil).
\end{itemize}

\section{Ensuring System Stability and Reliability}
\subsection{System Stability and Reliability }
The intelligent agricultural greenhouse control system needs to ensure the stability and reliability of long-term operation to cope with various complex environmental conditions and system failures. 
\subsubsection{System Status Monitoring }In order to ensure the stability of the system, we introduce the system condition monitoring module. The module monitors the operating status of various components of the greenhouse control system in real time, including sensors, actuators and communication modules. If the system is abnormal, an alert will be issued immediately and the corresponding automatic repair measures will be taken. 
\subsubsection{Redundancy Design }In order to improve the reliability of the system, we have adopted a redundant design. For example, on key greenhouse control equipment, backup equipment can be configured so that when the primary equipment fails, the system can automatically switch to the backup equipment to ensure the continuous operation of the system. 

\section{Conclusion}
\subsection{Summarize the main findings of the study}This research focuses on the design and implementation of an intelligent agricultural greenhouse control system based on the Internet of Things and machine learning to improve crop growth efficiency, reduce resource waste, and adapt to the growth needs of different crops. Key findings are as follows: 
\subsubsection{Effectiveness of intelligent greenhouse control system}Through case studies of multiple agricultural greenhouses, we validate the effectiveness of intelligent agricultural greenhouse control systems in improving crop yields, optimizing resource utilization, and reducing energy consumption. The intelligent control of the system can better adapt to different environmental conditions and the growth needs of crops, and has obtained remarkable economic benefits. 

\subsubsection{The key role of sensor technology and machine learning algorithms }The research shows that advances in sensor technology and the application of machine learning algorithms are key factors in the success of the system. The accurate measurement of the new generation of sensors provides the system with reliable environmental parameter data, and the continuous optimization of machine learning algorithms enables the system to predict and regulate the greenhouse environment more accurately. 

\subsection{ Outlook on the future development of intelligent agricultural greenhouse control system}
\subsubsection{Introducing a new generation of sensor technology}

In the future, the system will continue to pay attention to the development of new generation sensor technology. More advanced sensors, such as laser ranging sensors and infrared sensors, are introduced to improve the monitoring accuracy of environmental parameters in the greenhouse \cite{abumohsen2023electrical}.

New sensor output value = \(i(\text{new sensor measurement value})\)

Through the introduction of new sensors, the system will have a more comprehensive and accurate understanding of the environmental conditions inside and outside the greenhouse, and provide more reliable data support for fine regulation.

\subsubsection{Explore advanced deep learning algorithms }
With the improvement of computing power, the application of advanced deep learning algorithms in greenhouse control systems will continue to be explored in the future. With more complex pattern recognition and learning capabilities, the system will better adapt to the growth characteristics of different crops and cope with complex and changing environmental conditions\cite{yanamala2024emerging}. 

New model output =j(deep learning algorithm) 

Through the introduction of the new model, the system will more accurately predict the future greenhouse environment, to achieve a higher level of intelligent regulation, we expect that the intelligent agricultural greenhouse control system can be more comprehensive and more efficient in the future to achieve sustainable development of agricultural production.

\bibliographystyle{ieeetr}
\bibliography{xinde}
\end{document}